\documentstyle[preprint,tighten,aps]{revtex}
\begin{document}

\preprint{Submitted to Phys. Rev. Lett.}
\title{Far Dissipation Range of Turbulence}

\author{Shiyi Chen and Gary Doolen}
\address{Theoretical Division and Center for Nonlinear Studies\\
Los Alamos National Laboratory, Los Alamos, New Mexico 87545}

\author{Jackson R. Herring}
\address{National Center for Atmospheric Research, Boulder, Colorado 80302}

\author{Robert H. Kraichnan}
\address{369 Montezuma 108, Santa Fe, New Mexico 87501}

\author{Steven A. Orszag}
\address{Program in Applied and Computational Mathematics\\
Princeton University, Princeton, New Jersey 08544}

\author{Zhen Su She}
\address{Mathematics Department, University of Arizona, Tucson, Arizona 85721
\vskip .2truein}

\date{January 25, 1993}

\maketitle

\begin{abstract}
The very small scales of isotropic, Navier-Stokes turbulence at Reynolds number
${\cal R}_\lambda \approx 15$ are studied by high-resolution direct numerical
simulation (DNS) and by integration of the direct-interaction (DIA) equations.
The DNS follows the tail of the energy spectrum over more than thirty decades
of magnitude. The energy spectrum in the far-dissipation range $5k_d < k <
10k_d$ is well-fitted by $k^\alpha\exp(-ck/k_d)$, where $k_d$ is the Kolmogorov
dissipation wavenumber, $\alpha \approx 3.3$ and $c\approx 7.1$. For values of
$m$ that emphasize the far-dissipation range, the fields $(-\nabla^2)^m{\bf u}$
exhibit strong spatial intermittency, associated with gentle spatial variations
of the lower-$k$ part of the velocity field. DIA analysis gives a prefactor
$k^3$ and an exponential decay more rapid than DNS. Averaging over an ensemble
of DIA solutions, suggested by the observed intermittency, removes some of the
discrepancy.
\end{abstract}
\eject

\narrowtext

The smallest scales of incompressible, isotropic Navier-Stokes turbulence are
associated with the far-dissipation range of wavenumbers $k\gg k_d$, where $k_d
= \epsilon^{1/4}\nu^{-3/4}$ is the Kolmogorov dissipation wavenumber,
$\epsilon$ is the rate of dissipation of hydrodynamic kinetic energy per unit
mass and $\nu$ is kinematic viscosity. Most of the dissipation takes place at
$k<k_d$. The wavenumbers $k\gg k_d$ have attracted attention for a number of
years. There has been controversy concerning the asymptotic form of the energy
spectrum as $k\to\infty$. The smallest scales are of further interest because
they display strong intermittency even at Reynolds numbers so low that there is
no basis for a fractal cascade.

It is reasonable to assume that the wavenumbers $k\gg k_d$ represent spectral
tails of flow structures of spatial scale $\ge 1/k_d$. An analogy is the
exponential spectral tail of a shock that obeys Burgers' equation. There are
other possibilities. One is that very-high-wavenumber excitation comes mostly
from exceptionally strongly strained regions that give rise to observed
exponential-like skirts of the probability distribution function (pdf) of
vorticity.

A number of authors have discussed kinetic energy spectra for $k\gg k_d$ of the
form
\begin{equation}
E(k) \propto f(k/k_d)\exp[-c(k/k_d)^n]                           \label{eq1}
\end{equation}
where $c$ is a constant, $f$ is a weak function of $k/k_d$ and $1\le n\le 2$
\cite{1,2,3,4,5,6,7}. The direct interaction approximation (DIA), a
perturbative treatment, gives $n=1$ and $f \propto (k/k_d)^3$ \cite{2}.
Perturbation approximation can be justified for $k\gg k_d$ because the mode
amplitudes are very nearly the linear response, under molecular viscosity, to
quadratic forcing by modes of lower $k$ \cite{8}. The additional, unjustified
assumption in the DIA analysis is that the statistics of wave-vector amplitudes
for $k\gg k_d$ are nearly Gaussian. The plausible effect of the intermittency
actually present in the far-dissipation range is to enhance mean nonlinear
transfer and thereby raise the spectrum level above that predicted by DIA.

Foias, Manley and Sirovich \cite{3} have shown that $n\ge1$, under certain
assumptions of smoothness of the velocity field in a finite box. In view of
this inequality, The DIA results suggest that $n=1$ is exact for a finite box.
The particular form
\begin{equation}
E(k) \propto k^\alpha\exp(-ck/k_d) \label{eq2}
\end{equation}
seems consistent with a body of experimental and computer data \cite{7}.

Strong intermittency in the far-dissipation range at modest Reynolds numbers
was predicted some years ago on the basis of a simple physical argument
\cite{8}: $E(k)$ falls off steeply for $k \gg k_d$. Consequently, fluctuation,
on spatial macroscales, of parameters like $k_d$ in (\ref{eq1}) yields spatial
intermittency at scales $O(1/k)$ that increases without limit as $k/k_d \to
\infty$.

An ongoing computation project has achieved resolutions up to $512^3$
(wavenumber range $1\le k\le 256$) on a CM-200 computer at Los Alamos National
Laboratory \cite{9}. The direct numerical simulation (DNS) described in this
paper is limited by arithmetic precision rather than its resolution of $256^3$
(wavenumber range $1 \le k\le 128$). A nominal steady state was maintained by
forcing confined to $k < 3$, at a level determined to give the desired value of
the Taylor microscale Reynolds number ${\cal R}_\lambda$. The Taylor microscale
$\lambda$ for an isotropic turbulent flow is a length defined by
$\lambda=(15\nu v_0^2/\epsilon)^{1/2}$, where $v_0^2$ is the mean-square
velocity in any direction, and ${\cal R}_\lambda \equiv v_0\lambda/\nu$.

The solid line in Fig.~\ref{fig1} represents the time-averaged wavenumber
spectrum of kinetic energy $E(k)$ for a run with $\nu=0.026$, ${\cal R}_\lambda
\approx 14.9$, $k_d \approx 9.65$. The very close fit of the high-$k$ part of
the spectrum to a straight line (exponential decay) is apparent.
Fig.~\ref{fig2} shows $k\,d\ln E(k)/dk$ vs $k$ for this data. If $E(k)$ has the
form (\ref{eq2}), this plot is a straight line whose slope is $-c/k_d$ and
whose intercept on the vertical axis is $\alpha$. The straight line in
Fig.~\ref{fig2} is a least-squares fit to the data over the range $50 \le k \le
100$; it gives $c \approx 7.1$, $\alpha \approx 3.3$. Equation\ (\ref{eq2})
seems well supported. The confidence level for $\alpha$ is not high, because it
is not certain that the wavenumber range is long enough to give strictly
asymptotic results. It cannot even be asserted that $f(k/k_d)$ is exactly a
power. However, the value $n=1$ in (\ref{eq1}) does seem strongly favored.

A previous study \cite{7} gave a negative value for $\alpha$. The range used
for fitting in \cite{7} was $0.5k_d \le k \le 3k_d$, which is too low to give
asymptotic behavior. Note that in Fig.~\ref{fig2} the data points curve
downward at small $k$. An intercept $\alpha=0$ corresponds to tangency at $k=30
\approx 3k_d$ \cite{10}.

The dotted lines in Fig.~\ref{fig1} show the spectra in three subregions of the
cyclic box, defined by the $x$-space filter $\exp[-|{\bf x}-{\bf
x}_c|^2/(L/32)^2]$, where $L=2\pi$ is the box size and ${\bf x}_c$ is the
subregion center. The nominal linear dimension of a subregion is thus $L/16$.
In order to sufficiently reduce errors from chopping, the cyclic box was
repeated in each direction to give a total of $3^3$ replicas before the
filtering. The filtered field was transformed to $k$-space and subjected to
solenoidal projection before the spectrum was computed. An effect of the
filtering is marked depression of spectrum level for $k\lesssim 16$.

We have found that the distribution of spectral slopes over a set of 64
subregions, evenly spaced in the cyclic box, is consistent with the picture of
intermittency \cite{8} in which the parameters describing the spectral tail are
slowly-varying functions of spatial position. The three subregion spectra
plotted in Fig.~\ref{fig1} are those with minimum, median and maximum values of
$E(k=60)$. Note that the maximum and minimum values of $E(k=60)$ differ by a
factor of over $10^{10}$, despite the relatively small difference in slope.

Fig.~\ref{fig3} shows pdf's of the differentiated velocity fields
$(-\nabla^2)^m u_i$, averaged over the three components $i=1,2,3$. Note the
marked increase of intermittency with $m$. Fig.~\ref{fig4} is a visualization
of the field $\nabla^8u_1$. The shaded surfaces are where the absolute value of
field amplitude equals twice its root-mean-square value. The evident
three-dimensionality of these regions suggests intermittency that is associated
with gentle spatial variation on principal dissipation scales rather than with
exceptional regions that are strongly strained into thin sheets or tubes. The
flatness $\langle(\nabla^8u_1)^4\rangle /\langle(\nabla^8u_1)^2\rangle^2$ is
57. This large value is associated with the sharp peak of the pdf at zero
amplitude, rather than with the broad skirts at large amplitude values. The
latter represent probabilities too small to affect low-order statistics.

The spectral support of the field $\nabla^8u_1$ is effectively confined to the
range $15<k<40$. A field with spectral support confined to $50<k<100$, the
region where $E(k)$ is accurately proportional to $k^\alpha\exp(-ck/k_d)$, can
be constructed by applying the filter $\exp[-(k-75)^2/200]$ to the wave-vector
transform of $u_1({\bf x})$ and transforming back to $x$ space. Fig.~\ref{fig5}
shows the regions where the absolute value of the amplitude of this field
exceeds twice its root-mean-square value. The field $\nabla^{24}u_1$ has
approximately the same spectral support, but it is too noisy to give clean
visualizations.

The similarity between Figs.~\ref{fig4} and \ref{fig5} is striking. The regions
of high intensity are in the same locations and have similar shape, but are
smaller in Fig.~\ref{fig5}. This behavior suggests that the spectral support of
the very small scales represents the spectral tail of larger structures.
Similar behavior is exhibited by repeated differentiation of the velocity
profile of a Burgers shock.

We have integrated the DIA equations with the same forcing and viscosity as in
the DNS. The dashed line in Fig.~\ref{fig1} shows the result for $E(k)$. The
DIA spectrum tail falls within the range of DNS subregion values, but below the
median. There are two obvious causes for discrepancy between DIA and DNS. One
is the depression of high-$k$ energy transfer in DIA by sweeping effects in
response functions \cite{2}. At $k=75$, the sweeping decorrelation frequency
$v_0 k$ is about 30\% of the viscous decay frequency $\nu k^2$. Two-point
closures that are invariant to random Galilean transformation do not display
the strong sweeping decorrelation at large $k$ \cite{11}.

The second cause is the observed intermittency, which is not captured by DIA.
Consider the mean spectrum obtained by averaging over a finite-size ensemble of
DIA solutions, or other closure solutions, with different values of spectrum
level in the forcing region. As $k \to \infty$, this mean spectrum is dominated
by the ensemble member with the smallest value of $c/k_d$. Thus the effective
value of $c/k_d$ is decreased by the averaging. The prefactor exponent $\alpha
= 3$ is unchanged by the averaging. This latter value is a property of a broad
family of two-point closures \cite{7}. The averaging over closure predictions
is not explored further here, but it remains possible that simple statistical
distributions of the input parameters to the closure can give satisfactory
approximations both to $E(k)$ and to the large flatness factors of the fields
$(-\nabla^2)^m{\bf u}$.

The procedure of averaging over solutions with different parameter values has
an analytical implication that does not depend on closure approximation.
Suppose that the box containing the flow is of infinite size, but with spectral
support confined to wavenumbers above some wavenumber $k_0$. A local
wave-vector analysis centered on an arbitrary point ${\bf x}_c$ can be carried
out by applying the filter $\exp(-k_0^2|{\bf x}-{\bf x}_c|^2)$ to the velocity
field and performing solenoidal projection on the result.

Suppose that the spectral tail thereby associated with a region of the box with
dimensions $O(1/k_0)$ has the asymptotic behavior (\ref{eq2}), where $k_d$ is
now a weak function of ${\bf x}_c$. If any value of $k_d({\bf x}_c)$, however
large, occurs for some ${\bf x}_c$, then averaging over the entire infinite box
will give something slower than exponential decay of the spectrum as
$k\to\infty$. For example, if the pdf of $k_d$ over the box were $\propto
\exp(-k_d^2/k_c^2)$ as $k_d\to\infty$, where $k_c$ is a constant parameter,
then averaging of (\ref{eq2}) over this pdf would give a full-box spectrum
$E(k) \propto \exp[-(2ck/k_c)^{2/3}]$ as $k\to\infty$, apart from an algebraic
prefactor. Finally, if the size of the box were made finite but $\gg 1/k_0$,
then this full-box spectrum would change over from stretched-exponential to
simple-exponential decay at some wavenumber $\gg k_c$. The bound obtained by
Foias, Manley and Sirovich \cite{3} assumes a finite box.

We thank Xiaowen Shan for help with the design of the computations and
acknowledge valuable discussions with S.~Kida, O.~Manley, R.~S. Rogallo,
F.~Waleffe, and Y.~Zhou. D.~W. Grunau and C.~D. Hansen kindly produced the
visualizations. This work was supported by the Department of Energy, the
National Science Foundation, the Defense Advanced Research Projects Agency, and
the Office of Naval Research. The computations were performed at the Advanced
Computing Laboratory, Los Alamos National Laboratory and at the National Center
for Atmospheric Research.

\vskip .5truein\section*{\large FIGURE CAPTIONS}\firstfigfalse

\begin{figure}
\caption{\label{fig1}
Linear-log plot of $E(k)$ vs. $k$. Solid line, DNS spectrum; dotted lines,
subregional spectra with largest, median and smallest values of $E(k=60)$;
dashed line, DIA spectrum.}
\end{figure}

\begin{figure}
\caption{\label{fig2}
The function $k\,d\ln E(k)/dk$ vs $k$ for the DNS spectrum. The straight line
is a least-squares fit to the data points for $50\le k\le100$.}
\end{figure}

\begin{figure}
\caption{\label{fig3}
Pdf $P(w)$ of the field $w = (-\nabla^2)^m u_i$ for $m=0$ (dashed), $m=2$
(dotted) and $m=4$ (solid), averaged over $i=1,2,3$.}
\end{figure}

\begin{figure}
\caption{\label{fig4}
Perspective view of the surface where $|\nabla^8u_1|$ equals twice its
root-mean-square value.}
\end{figure}

\begin{figure}
\caption{\label{fig5}
Perspective view of the surface where the absolute value of the filtered field
with spectral support centered at $k=75$ equals twice its root-mean-square
value.}
\end{figure}

\end{document}